\documentclass[11pt]{article}

\usepackage[fleqn]{amsmath}
\usepackage{amsfonts}
\usepackage{amssymb}
\usepackage{epsfig}

\newcommand{\drm}{\mathrm{d}}

\newcommand{\pddt}{\frac{\partial\phantom{t}}{\partial t}}

\newcommand{\refeq}[1]{(\ref{#1})}

\newcommand{\vect}[1] {\boldsymbol{{ #1}} }

\newcommand{\Rset}{\mathbb{R}}

\newcommand{\jV}{{\vect{j}}}		% 3-current density
\newcommand{\sV}{{\vect{s}}}            % position 3-vector
\newcommand{\AV}{\pmb{{\cal A}}}
\newcommand{\BV}{\pmb{{\cal B}}}
\newcommand{\DV}{\pmb{{\cal D}}}
\newcommand{\EV}{\pmb{{\cal E}}}

\newcommand{\HV}{\pmb{{\cal H}}}

\newcommand{\nab}{\vect{\nabla}}

\renewcommand{\geq}{\geqslant}

\reversemarginpar

\numberwithin{equation}{section}

\begin{document}

%%%%%%%%%%%%%%%%%%%%%%%%%%%%%%%%%%%%%%%%%%%%%%%%%%%%%%%%%%%%%%%%%%%%%
%%%%%%%%%%%%%%%%%%%%%%%%%%%%%%%%%%%%%%%%%%%%%%%%%%%%%%%%%%%%%%%%%%%%%

\title{\uppercase{Some uniqueness results for stationary solutions to the \\
		Maxwell-Born-Infeld field equations \\
		and their physical consequences}}

\author{\textbf{Michael K.-H. KIESSLING}\\
                Department of Mathematics, Rutgers University\\
                110 Frelinghuysen Rd., Piscataway, NJ 08854, USA\\ 
		miki@math.rutgers.edu \\ \\ \\
\textrm{Revised version, August 27, 2011}\\
\textrm{This printout was typeset with \LaTeX\ on}}
\maketitle
\thispagestyle{empty}

\begin{abstract}
{\noindent 
  Uniqueness is established for time-independent finite-energy electromagnetic fields which solve 
the nonlinear Maxwell--Born--Infeld equations in boundary-free space under the condition that either 
the charge or current density vanishes. 
 In addition, it is also shown that the simpler Maxwell--Born equations admit at most one
stationary finite-energy electromagnetic field solution, without the above condition.
 In these theories of electromagnetism the following physical consequences emerge:
source-free field solitons moving at speeds less than the vacuum speed of light $c$ do not exist;
any electrostatic (resp. magnetostatic) field is the unique stationary electromagnetic 
field for the same current-density-free (resp. charge-density-free) sources.
 These results put to rest some interesting speculations in the physics literature.
}
\end{abstract}

\vfill\vfill
\hrule
\smallskip
\copyright(2011) \small{The author. Reproduction of this preprint, in its entirety, is 

permitted for non-commercial purposes only.}

\newpage

%%%%%%%%%%%%%%%%%%%%%%%%%%%%%%%%%%%%%%%%%%%%%%%%%%%%%%%%%%%%%%%%%%%%%%%%%%%

\baselineskip=14pt

%%%%%%%%%%%%%%%%%%%%%%%%%%%%%%%%%%%%%%%%%%%%%%%%%%%%%%%%%%%%%%
%%%%%%%%%%%%%%%%%%%%%%%%%%%%%%%%%%%%%%%%%%%%%%%%%%%%%%%%%%%%%%
                \section{Introduction}
%%%%%%%%%%%%%%%%%%%%%%%%%%%%%%%%%%%%%%%%%%%%%%%%%%%%%%%%%%%%%%
%%%%%%%%%%%%%%%%%%%%%%%%%%%%%%%%%%%%%%%%%%%%%%%%%%%%%%%%%%%%%%

	Frustrated by the ultraviolet divergences of pre-renormalization QED, 
in the early 1930s Max Born \cite{BornA} inaugurated an intriguing alternate 
quest for a quantum theory of electromagnetism.
        He argued that we (physicists) were trying to quantize the wrong set of classical 
electromagnetic field equations, and that the correct one would have to be a \emph{nonlinear} realization 
of {Maxwell's electromagnetic field equations} in space ($\sV$) and time ($t$). 
	A nonlinear classical electromagnetic field theory program was already available since 
the pioneering work of Gustav Mie \cite{MieFELDTHEORIE}, but it didn't provide a physical criterion
that would single out a particular nonlinearity.
	By an act of serendipity \cite{BiBiONE}, Born and Infeld \cite{BornInfeldB} then proposed a distinct
one-parameter family of field equations, the distinction of which was noticed only much later, by
Boillat \cite{Boillat} and Plebanski \cite{Plebanski}.
       Namely, these equations emerge as the \emph{unique} set 
of electromagnetic field equations derivable from a Lagrangian which:
\smallskip

(L) are covariant under the Lorentz\footnote{Actually they satisfy (P): 
		covariance under the larger Poincar\'e group.}
 group;

(W) are covariant under the gauge (i.e. Weyl) group;

(M) reduce to the linear Maxwell field equations in the weak field limit;

(E) have finite field-energy solutions with point-charge sources;

(D) are linearly completely degenerate.\footnote{While the unique 
		characterization of these Maxwell--Born--Infeld field equations 
		in terms of (L),(W),(M),(E),(D) was only discovered in 1970, the fact that these 
		field equations do satisfy, beside Born--Infeld's (L),(W),(M),(E), also (D) was 
		already known to Schr\"odinger as \emph{absence of birefringence} (double refraction), 
		see p.102 in \cite{ErwinDUBLINa}.}

\smallskip
\noindent
        A further distinction, in terms of a mildest-singularity criterion for curved spacetimes, was 
recently discovered by Tahvildar-Zadeh \cite{Shadi}.

	The uniqueness of the Maxwell--Born--Infeld field equations under these five 
\emph{physically very reasonable} postulates provides a compelling argument for their study. 
	In the weak field regime they inevitably reproduce all the well-known 
electromagnetic phenomema that agree with solutions of the conventional linear 
Maxwell field equations ``in vacuo,'' which satisfy (L),(W),(M), and (D), but not (E).
	Since (L),(W),(M), and (E) must be the backbone of \emph{any} nonlinear classical theory of 
electromagnetism which does not suffer from infinite self-energies of point charges, it is currently 
an active field of inquiry to test the validity of (D) in the ``classical strong-field regime'' of 
large field strengths which do not yet require quantum-physical considerations \cite{AielloETal, BurtonETal}.
	This is equivalent to finding out whether the Maxwell--Born--Infeld equations make also the correct 
strong-field predictions.

	The Maxwell--Born--Infeld field equations comprise Maxwell's evolution equations for the 
magnetic induction field $\BV$ and the electric displacement field $\DV$, 
\begin{alignat}{1}
\textstyle
   \frac{1}{c}
\pddt{\BV(t,\sV)}
&= \label{eq:MdotB}
        - \nab\times\EV(t,\sV) \, ,
\\
\textstyle
        \frac{1}{c}\pddt{\DV(t,\sV)}
&= 
        + \nab\times\HV(t,\sV)  - 4\pi \textstyle{\frac{1}{c}} \jV(t,\sV) \, ,
\label{eq:MdotD}
\end{alignat}
constrained by the two scalar equations 
\begin{alignat}{1}
        \nab\cdot \BV(t,\sV)  
&= \label{eq:MdivB}
        0\, ,
\\
        \nab\cdot\DV(t,\sV)  
&=
        4 \pi \rho (t,\sV) \, .
\label{eq:MdivD}
\end{alignat}
	These Maxwell field equations are understood in the sense of distributions.
	The nonlinearity comes in through a nonlinear relationship amongst the four fields 
$\EV,\BV,\DV,\HV$, replacing Maxwell's linear ``law of the pure aether:'' $\EV=\DV$ and $\HV=\BV$.
	The nonlinearity proposed by Born and Infeld \cite{BornInfeldB} reads
\begin{alignat}{1}
&{\EV} 
= \label{eq:FOLIeqEofBD}
\frac{{\DV}-\frac{1}{b^2}{\BV}\times({\BV}\times{\DV})}
     {\sqrt{ 1 + \frac{1}{b^2}(|{\BV}|^2 + |{\DV}|^2) + \frac{1}{b^4}|{\BV}\times {\DV}|^2}}\,,
\\
&{\HV} 
 = \label{eq:FOLIeqHofBD}
\frac{{\BV}-\frac{1}{b^2}{\DV}\times({\DV}\times {\BV})}
     {\sqrt{ 1 + \frac{1}{b^2}(|{\BV}|^2 + |{\DV}|^2) + \frac{1}{b^4}|{\BV}\times {\DV}|^2}}
\, ,
\end{alignat}
where $b\in (0,\infty)$ is \emph{Born's field strength}, a hypothetical new ``constant of nature.'' 
	By dropping the $\BV\times\DV$ terms one obtains the ``Maxwell--Born field equations.'' 

	The charge density $\rho (t,\sV)$ and electric current vector-density $\jV (t,\sV)$,
which jointly satisfy (in the sense of distributions) the local law of charge conservation 
\begin{equation}
\textstyle
        \pddt\rho (t,\sV) + \nab\cdot\jV(t,\sV)  
=\label{eq:MrhojLAW}
	0,
\end{equation}
are to be supplied by a relativistic theory of matter.
	Originally Born \cite{BornA} seems to have had in mind only moving point charge source terms, 
representing point electrons, though subsequently he also contemplated electrons as given by a
spinning charged ring singularity, see \cite{RaoRING}. 
	We will allow somewhat more general sources.

	Unfortunately, since their nonlinearity is quite formidable, our knowledge of the solution 
properties of the Maxwell--Born--Infeld field equations beyond the weak-field approximation is still quite 
rudimentary and fragmentary.
	  In particular, neither exact special solutions nor rigorous general results seem to be available
for fields with dynamical sources.
	Closed-form expressions for special solutions and rigorous general results about solutions 
have been obtained only for field evolutions in the absence of sources, and for time-independent fields 
with stationary sources.

     More specifically, after a false start by others \cite{ChaeHuh}, Speck \cite{SpeckMBI} recently proved
the important result that the classical initial-value problem for the source-free Maxwell--Born--Infeld field 
equations is globally well-posed at least for sufficiently small finite-energy initial data prescribed
on a spacelike Cauchy hypersurface in Minkowski spacetime; for the hyperbolicity of the equations,
see also \cite{Brenier,SerreC,Perlick}.
   The small data constraint is presumably necessary because the formation of singularities in 
finite time is expected for large data, but so far nobody seems to have a genuinely three-dimensional 
blow-up result. 
  Much more is known about source-free  plane-wave solutions \cite{ErwinDUBLINa, AielloETal, ErwinDUBLINc}.
  In that case, both global existence and finite-time blow-up results are available, see 
\cite{Brenier, SerreA, SpeckTHESIS}, but 
it is not clear how representative of genuinely three-dimensional field evolutions some of 
the intriguing features of such plane-wave solutions really are.

  In the presence of time-independent charge and current densities $\rho$ and $\jV$ the following is 
known about stationary finite-energy solutions in entire space  (i.e., no boundaries).
  If integrable $\rho$ and $\jV$ are sufficiently \emph{small and regular} (in some H\"older space), then
the stationary Maxwell--Born--Infeld field equations have a classical electromagnetic solution
which can be computed with an absolutely convergent perturbative series in inverse powers of Born's 
parameter $b$ \cite{CarKieSIAM,KieMBIinJMP}; and when either $\rho$ or $\jV$ vanish identically, these solutions 
are purely magnetostatic or purely electrostatic, respectively, and then also unique within these
respective subclasses of solutions.
   It is not clear whether the power series also converges, in some generalized sense, to a solution 
for large regular or for singular point or ring sources. 
   Yet it can be shown \cite{KieMBIinCMPpreprint} with a variational argument  that a unique 
electrostatic finite-energy solution exists for each arbitrary choice of finitely many point 
charge sources (with arbitrary locations, signs and magnitude of charge),
and elliptic regularity theory shows that these solutions are real analytic 
away from the locations of the point charge; see also earlier, more restricted results in \cite{KlyMik, Kly}.
  Known in closed form so far are only the electrostatic solutions for a 
single point charge \cite{BornA,BornInfeldA,Ecker,GibbonsA} and for an infinite crystal \cite{HoppeA};
some of these have infinite energy, though.

	In this communication we will add some basic uniqueness results which, to the 
best of our knowledge, are new.
	Their precise statements and proofs will be given in sections 3 and 5.
	Here we summarize those of section 3 as follows: 

\noindent
\emph{Whenever one of the stationary sources $\rho$ or $\jV$ vanishes identically, then any
time-independent finite-energy solution of the Maxwell--Born--Infeld field equations in the 
entire physical space is unique.
 Moreover, this unique solution is purely electrostatic in case $\jV\equiv \vect{0}$ and purely magnetostatic 
in case $\rho\equiv 0$.}

        We emphasize that our results say nothing about the uniqueness of stationary solutions to the
Maxwell--Born--Infeld equations when both $\rho$ and $\jV$ do not vanish identically.  
        To sort this out is an interesting challenge.
	Meanwhile, for the Maxwell-Born field equations we can say more.
	We summarize our results of section 5 as:

\noindent
\emph{Any stationary finite-energy electromagnetic field in entire space which solves
the Maxwell--Born equations, with both $\rho$ and $\jV$ allowed to be nontrivial, is unique.}
\newpage

	Lest the reader thinks that uniqueness results are only of interest to mathe\-matically-minded 
readers, we hasten to emphasize that our results put to rest  speculations in the physics literature as to
whether the Maxwell--Born--Infeld equations predict such interesting types of phenomena as field solitons 
and electromagnetic self-induction. 
	This is explained in section 4.
	But first, after introducing (in section 2) two basic field functionals, 
the Lagrangian and its Legendre transform w.r.t. $\EV$, the Hamiltonian, which serves to define what is 
meant by a ``finite-energy solution,'' we state and then rigorously prove our uniqueness results for the
Maxwell--Born--Infeld field equations (section 3).
	Finally, in section 5 we state and then prove our uniqueness result for the Maxwell--Born field equations
and conclude with an outlook on extensions of our results to other nonlinear electromagnetic field theories.

%%%%%%%%%%%%%%%%%%%%%%%%%%%%%%%%%%%%%%%%%%%%%%%%%%%%%%%%%%%%%%
%%%%%%%%%%%%%%%%%%%%%%%%%%%%%%%%%%%%%%%%%%%%%%%%%%%%%%%%%%%%%%
                \section{The Lagrangian and the Hamiltonian}
%%%%%%%%%%%%%%%%%%%%%%%%%%%%%%%%%%%%%%%%%%%%%%%%%%%%%%%%%%%%%%
%%%%%%%%%%%%%%%%%%%%%%%%%%%%%%%%%%%%%%%%%%%%%%%%%%%%%%%%%%%%%%

	The Maxwell--Born--Infeld field equations with general (prescribed) sources derive from a Lorentz
and Weyl covariant action principle. 
	For stationary situations, the time integrations are trivial and one is left with a variational
principle for a space integral, the Lagrangian functional of the potential variables $\phi$ and $\AV$,
given by
\begin{equation} 
\hskip-5pt
  L (\phi,\AV)
= \label{eq:LAGRANGIAN}
\int_{\mathbb R^3}\!\left[
\textstyle{  \frac{b^2}{4\pi}}\Bigl(1-\!{\sqrt{1 - \textstyle{\frac{1}{b^2}}(|{\EV}|^2 - |{\BV}|^2) 
			              - \textstyle{\frac{1}{b^4}}|{\EV}\cdot {\BV}|^2}}\Bigr)
- \rho \phi + {\textstyle\frac{1}{c}}\jV\cdot\AV\right]\drm^3s
\end{equation}
where $\BV=\nab\times\AV$ and $\EV=-\nab\phi$.
	Although not manifestly obvious, it is nevertheless straightforward to show that $L (\phi,\AV)$
is convex in $\phi$ and concave in $\AV$. 
	Therefore, the stationary special case of the Maxwell--Born--Infeld field equations with 
prescribed stationary sources is obtained by seeking the minimum of $L$ w.r.t. $\phi$ and its maximum 
w.r.t. $\AV$ --- its critical points are saddle points in $\phi,\AV$ space.

	With the help of the stationary Maxwell--Born--Infeld field equations (with sources)
one can rewrite $L$ at any of its critical points, say $(\phi_{0},\AV_{0})$, in terms of 
the variables $(\DV_{0},\AV_{0})$. 
	Using Coulomb's law \refeq{eq:MdivD}, an integration by parts and the 
definition $\EV=-\nab\phi$, we find   
$L (\phi_{0},\AV_{0}) =  - H (\DV_{0},\AV_{0})$, where
\begin{equation} 
\hskip-5pt
 H (\DV,\AV)
= \label{eq:LAGRANGIANatCRITpt}
\int_{\mathbb R^3}\!\left(
\textstyle{\frac{b^2}{4\pi}}
\Bigl[\!{\sqrt{1 + \textstyle{\frac{1}{b^2}}(|{\BV}|^2 + |{\DV}|^2) 
	       + \textstyle{\frac{1}{b^4}}|{\BV}\times {\DV}|^2}} -1\Bigr]
- {\textstyle\frac{1}{c}}\jV\cdot\AV\right)\drm^3s,
\end{equation}
with $\BV=\nab\times\AV$.
 The \emph{stationary} field equations with stationary prescribed sources are recovered 
from the constrained energy principle of minimizing $H$ w.r.t. $\AV$ and $\DV$ 
\emph{under the constraint} \refeq{eq:MdivD}, and with $\BV=\nab\times\AV$.

 The integral 
\begin{equation}
\hskip-5pt
  E (\BV,\DV)
= \label{eq:FIELDenergyMBI}
{\textstyle{\frac{b^2}{4\pi}}}
\int_{\mathbb R^3}\!
\Bigl[\!{\sqrt{1 + \textstyle{\frac{1}{b^2}}(|{\BV}|^2 + |{\DV}|^2) 
	       + \textstyle{\frac{1}{b^4}}|{\BV}\times {\DV}|^2}} -1\Bigr] \drm^3s
\end{equation}
is called the field energy.
 It is characterized by a Legendre-Fenchel transform,
\begin{equation} 
\hskip-4.5pt
4\pi E (\BV,\DV)  
= \label{eq:LEGENDREtransf}
\max_{\EV}\! \int_{\mathbb R^3}\!\left(
b^2
\Bigl[\!{\sqrt{1 - \textstyle{\frac{1}{b^2}}(|{\EV}|^2 - |{\BV}|^2) 
	       - \textstyle{\frac{1}{b^4}}|{\EV}\cdot {\BV}|^2}}-1\Bigr]\! + \DV \cdot\EV\right)\drm^3s.
\end{equation}
 As a result, $E(\BV,\DV)$ is strictly convex in $\DV$, and by its $\BV\leftrightarrow\DV$ symmetry, 
it is also strictly convex in $\BV$. 
 Unfortunately,  though, $E(\BV,\DV)$ is not \emph{jointly} convex in $\BV$ and $\DV$, as the following 
counterexample shows: the Hessian of the integrand of $E(\BV,\DV)$ is a $6\times6$-matrix function 
on $\BV,\DV$ space, and for the trial vectors $\BV_*=(1,2,3)$ and $\DV_*=(4,5,6)$ it has 4 positive 
and 2 negative eigenvalues. 
 It is now straightforward to show that the second variation of $E(\BV,\DV)$ is indefinite.
 Indeed, just extend $\BV_*$ and $\DV_*$ continuously to vector fields on $\Rset^3$ 
which are the constant vectors (1,2,3) and (4,5,6) over a vast domain $D$, then suitably 
decay rapidly to zero; taking variation vector fields which are identical to an eigenvector for one of the 
negative (positive) eigenvalues of the Hessian over $D$ yields a negative (positive) second variation
because the contributions from outside $D$ can be made arbitrarily small relative to those from $D$.

 The functional $H$ inherits these convexity properties, expressed in $\DV$ and $\AV$.
 Its strict convexity in $\BV$ and $\DV$ will play an important role in our uniqueness proofs, 
but the failure of joint convexity is the reason for why we do not have a uniqueness result when 
both $\rho$ and $\jV$ are nonvanishing.
 This does not yet mean that uniqueness fails when both $\rho$ and $\jV$ are nonvanishing; our
proof technique fails, though.

 As a side remark we note that both $H$ and $E$ can serve as Hamiltonians for the dynamical fields, 
the former generally, with $\DV$ and $\AV$ as canonical pair, and the latter for source-free situations, 
when $\DV$ and $\BV$ can be used as canonical pair.

%%%%%%%%%%%%%%%%%%%%%%%%%%%%%%%%%%%%%%%%%%%%%%%%%%%%%%%%%%%%%%
%%%%%%%%%%%%%%%%%%%%%%%%%%%%%%%%%%%%%%%%%%%%%%%%%%%%%%%%%%%%%%
                \section{Uniqueness results for the Maxwell--Born--Infeld field equations}
%%%%%%%%%%%%%%%%%%%%%%%%%%%%%%%%%%%%%%%%%%%%%%%%%%%%%%%%%%%%%%
%%%%%%%%%%%%%%%%%%%%%%%%%%%%%%%%%%%%%%%%%%%%%%%%%%%%%%%%%%%%%%

	We will be quite general and allow a combination of finitely many Dirac (point and ring) sources 
with integrable (H\"older-)regular sources.
	As a result, any putative distributional field solution should be H\"older-continuously differentiable 
away from the Dirac sources; this has already been proven true in the purely electrostatic case 
\cite{CarKieSIAM,KieMBIinCMPpreprint} and for small H\"older sources also in the electromagnetic 
case \cite{KieMBIinJMP}.
	Let us stipulate once and for all that we are only discussing distributional solutions in all of $\Rset^3$
which are equivalent to such  regular functions.

	We now first state and then prove our main uniqueness results one by one.
	The proofs of all our uniqueness theorems are based on the strategy that the hypothesis of 
two distinct stationary solutions leads to an absurd conclusion.

	Our first theorem concerns source-free fields and, hence, is implied by
each of our subsequent Theorems 2 and 3.
	Yet, Theorem 1 is of independent interest, and its proof much simpler than those of the other theorems.
\smallskip

\noindent{\textbf{Theorem 1}.}
\emph{The only time-independent finite-energy solution of the Maxwell--Born--Infeld field equations
\refeq{eq:MdotB}-\refeq{eq:FOLIeqHofBD}\! with both charge density $\rho$ and current vector-density $\jV$ 
vanishing identically, is the trivial electromagnetic field $(\BV_0,\DV_0) = (\vect{0},\vect{0})$.}
\smallskip

\noindent
{\sc{Proof}}: Clearly, the trivial electromagnetic field $(\BV_0,\DV_0) = (\vect{0},\vect{0})$ solves the
Maxwell--Born--Infeld field equations \refeq{eq:MdotB}--\refeq{eq:FOLIeqHofBD} with identically
vanishing charge density $\rho$ and current vector-density $\jV$. 
  Now suppose that there exists another, nontrivial solution $(\BV_1,\DV_1)\not = (\vect{0},\vect{0})$.
  Then $(\BV_1,\DV_1)$ is a stationary point of $E(\BV,\DV)$, and so 
the first derivative w.r.t. $\lambda$ of $E(\lambda\BV_1,\lambda\DV_1)$ has to vanish at $\lambda=1$.
  But 
\begin{equation} 
\hskip-5pt
E (\lambda\BV_1,\lambda\DV_1)  
= \label{eq:EofBlDl}
{\textstyle{\frac{b^2}{4\pi}}}
\int_{\mathbb R^3}\!
\Bigl[\!{\sqrt{1 + \textstyle{\frac{\lambda^2}{b^2}}(|{\BV_1}|^2 + |{\DV_1}|^2) 
	       + \textstyle{\frac{\lambda^4}{b^4}}|{\BV_1}\times {\DV_1}|^2}} -1\Bigr] \drm^3s,
\end{equation}
and its $\lambda$-derivative at $\lambda=1$ is manifestly strictly positive. 
 Therefore, no nontrivial solution $(\BV_1,\DV_1)\not = (\vect{0},\vect{0})$ can exist.
\hfill QED.
\smallskip

	Our Theorem 1 extends to electromagnetic fields the results of \cite{Yisong} which cover
purely electrostatic or magnetostatic fields.
	Previously a genuinely electromagnetic version was known only under a
smallness condition on the field strengths \cite{KieMBIinJSPa}.

	Next we allow a nontrivial charge density $\rho$ while $\jV$ remains trivial.

\noindent{\textbf{Theorem 2}.}
\emph{Suppose the current vector-density $\jV$ vanishes identically. 
	Then the Maxwell--Born--Infeld equations \refeq{eq:MdotB}--\refeq{eq:FOLIeqHofBD}
have at most one time-independent finite-energy electromagnetic field solution for a given stationary 
charge density $\rho$, and this unique solution  $(\BV_0,\DV_0)$ is purely electrostatic, i.e.
$(\BV_0,\DV_0)=(\vect{0},\DV_0)$.}

\smallskip

\noindent
{\sc{Proof}}: Suppose there were two different stationary solutions with finite field 
energy, say $(\BV_{0},\DV_0)\not=(\BV_1,\DV_1)$.
	We also define the associated fields
\begin{alignat}{1}
{\HV}_k
 = \label{eq:HofBDone}
\frac{{\BV}_{k} -\frac{1}{b^2}{\DV}_{k}\times({\DV}_{k}\times {\BV}_{k})}
     {\sqrt{ 1 + \frac{1}{b^2}(|{\BV}_{k}|^2 + |{\DV}_{k}|^2) 
+ \frac{1}{b^4}|{\BV}_{k}\times {\DV}_{k}|^2}}
\,, \qquad k =0,1.
\end{alignat}
	Now, since both $(\BV_{0},\DV_0)$ and $(\BV_1,\DV_1)$ solve the stationary 
Maxwell--Born--Infeld field equations with the same $\rho$ and $\jV=\vect{0}$, we have
(for $k=0$ and 1) that
\begin{alignat}{1}
\textstyle
\nab\times\HV_k(\sV)  
&= 
 \vect{0} .
\label{eq:McurlHk}
\end{alignat}
	Furthermore, let $\AV_0$ and $\AV_1$ be magnetic vector potentials for $\BV_0$ and $\BV_1$,
respectively, defined in the sense of distributions; without loss of generality we may assume 
(but don't need to!) that they vanish as $|\sV|\to\infty$.
	We take the Euclidean inner product of \refeq{eq:McurlHk} with $\AV_k$, 
integrate over $\Rset^3$, use integration by parts to move the curl over 
to $\AV_k$ (for which we use that surface integrals at spatial infinity vanish thanks to 
the finite-energy condition), note that $\nab\times \AV_k =\BV_k$, and obtain
\begin{alignat}{1}
0 
= 
\int_{\mathbb R^3} \AV_k\cdot \nab\times \HV_k \drm^3s 
 =\label{eq:BdotHoneINTid}
\int_{\mathbb R^3} \BV_k\cdot \HV_k \drm^3s .
\end{alignat}
  But
\begin{alignat}{1}
\BV_k\cdot \HV_k
 = \label{eq:BdotHoneID}
\frac{|{\BV}_{k}|^2 +\frac{1}{b^2}|{\BV}_{k}\times {\DV}_{k}|^2}
     {\sqrt{ 1 + \frac{1}{b^2}(|{\BV}_{k}|^2 + |{\DV}_{k}|^2) 
+ \frac{1}{b^4}|{\BV}_{k}\times {\DV}_{k}|^2}}
 \, .
\end{alignat}
	Clearly, r.h.s.\refeq{eq:BdotHoneID} $\geq 0$, and ``$=$'' holds if and only if $\BV_k=\vect{0}$.
	So we have shown that any hypothetical solution is electrostatic: $(\BV_k,\DV_k)=(\vect{0},\DV_k)$.

	Yet there can only be a unique electrostatic finite-energy solution for a given $\rho$, as already
shown by Pryce \cite{PryceB}; cf. also \cite{GibbonsA} and \cite{KieMBIinJSPa}.
\hfill QED
\smallskip

	Our next theorem is a mirror image of Theorem 2.

\noindent{\textbf{Theorem 3}.}
\emph{Suppose the charge density $\rho$ vanishes identically. 
	Then the Maxwell--Born--Infeld equations \refeq{eq:MdotB}--\refeq{eq:FOLIeqHofBD}
have at most one time-independent finite-energy electromagnetic field solution for a given stationary 
current vector-density $\jV$, and this unique solution  $(\BV_0,\DV_0)$ is purely magnetostatic, i.e.
$(\BV_0,\DV_0)=(\BV_0,\vect{0})$.}
\smallskip

\noindent
{\sc{Proof}}: Not surprisingly, the proof of Theorem 3 is a mirror image of the proof of Theorem 2.
	So, suppose again there were two different stationary solutions with finite field 
energy, say $(\BV_0,\DV_0)\not=(\BV_1,\DV_1)$.
	We define the associated fields
\begin{alignat}{1}
{\EV}_k
 = \label{eq:EofBDone}
\frac{{\DV}_{k} -\frac{1}{b^2}{\BV}_{k}\times({\BV}_{k}\times {\DV}_{k})}
     {\sqrt{ 1 + \frac{1}{b^2}(|{\BV}_{k}|^2 + |{\DV}_{k}|^2) 
+ \frac{1}{b^4}|{\BV}_{k}\times {\DV}_{k}|^2}}
\, ,\qquad k=0,1.
\end{alignat}
	We let $\phi_k$ denote their electric potentials, viz. $\EV_k = -\nab \phi_k$, defined in the 
sense of distributions; without loss of generality we may assume (but again don't need to!) 
that they vanish as $|\sV|\to\infty$.

	Next, since both $(\BV_0,\DV_0)$ and $(\BV_1,\DV_1)$ solve the stationary 
Maxwell--Born--Infeld field equations with the same $\rho=0$ and $\jV$, we have (for $k=0$ and 1) that
\begin{alignat}{1}
        \nab\cdot\DV_k(\sV)  
&=
        0\, .
\label{eq:MdivDkNULL}
\end{alignat}
	We multiply \refeq{eq:MdivDkNULL}  by $\phi_k$, integrate over
$\Rset^3$, use integration by parts to move the $\nab$ operator over to $\phi_k$
(again using that surface integrals at spatial infinity vanish because of
the finite-energy condition), use $-\nab\phi_k =\EV_k$, and obtain
\begin{alignat}{1}
0 
= 
\int_{\mathbb R^3} \phi_k\nab\cdot\DV_k \drm^3s
 =
\int_{\mathbb R^3} \EV_k \cdot \DV_k \drm^3s .
\label{eq:DEoneISzero}
\end{alignat}
  But
\begin{alignat}{1}
\EV_k \cdot \DV_k 
 = 
\frac{|{\DV}_{k}|^2 +\frac{1}{b^2}|{\BV}_{k}\times {\DV}_{k}|^2}
     {\sqrt{ 1 + \frac{1}{b^2}(|{\BV}_{k}|^2 + |{\DV}_{k}|^2) 
+ \frac{1}{b^4}|{\BV}_{k}\times {\DV}_{k}|^2}}\, .
\label{eq:DoneDotEoneID}
\end{alignat}
	Clearly, r.h.s.\refeq{eq:DoneDotEoneID} $\geq 0$, and ``$=$'' holds if and only if $\DV_k=\vect{0}$.
	So we have shown that any hypothetical solution is magnetostatic:
$(\BV_k,\DV_k)=(\BV_k,\vect{0})$.

	Yet there can only be one magnetostatic finite-energy solution for a given $\jV$.
	The proof of this claim is the magnetostatic special case of the pertinent part of the uniqueness proof
for the Maxwell--Born equations, given in section 5.
\hfill QED
\smallskip
	
	We close this section by re-emphasizing that we have nothing to say about uniqueness of 
stationary electromagnetic solutions to the Maxwell--Born--Infeld field equations when both 
$\rho$ and $\jV$ are nonvanishing somewhere.

%%%%%%%%%%%%%%%%%%%%%%%%%%%%%%%%%%%%%%%%%%%%%%%%%%%%%%%%%%%%%%
%%%%%%%%%%%%%%%%%%%%%%%%%%%%%%%%%%%%%%%%%%%%%%%%%%%%%%%%%%%%%%
                \section{Physical implications for electromagnetism}
%%%%%%%%%%%%%%%%%%%%%%%%%%%%%%%%%%%%%%%%%%%%%%%%%%%%%%%%%%%%%%
%%%%%%%%%%%%%%%%%%%%%%%%%%%%%%%%%%%%%%%%%%%%%%%%%%%%%%%%%%%%%%

	In the conceivable event that the Maxwell--Born--Infeld field equations emerge
in the classical limit of the elusive divergence-problem-free quantum theory of
electromagnetism, the following physical conclusions hold for the classical realm:

	Our Theorem 1 bears on the search for source-free particle-like electromagnetic field 
structures, e.g. \cite{Benci}; remarkably, this search goes back at least to \cite{MieFELDTHEORIE}.
	Pauli \cite{Pauli} already noted that soliton-yielding electromagnetic field equations which
are semi-linear in the electromagnetic potentials usually run afoul of (W). 
	Pauli's criticism does not apply to the (W)-obeying  Maxwell--Born--Infeld equations which are quasi-linear 
in the potentials without a semi-linear part.
	However, our Theorem 1 implies:

\noindent
* \emph{Finite-energy source-free soliton-type electromagnetic field solutions in all of space 
traveling at speeds less than the vacuum speed of light $c$ do not exist.}

\noindent
	For suppose they would, then a Lorentz boost could be performed to their comoving frame.
	The boosted solution would be a non-trivial source-free stationary electromagnetic field of 
finite energy --- in contradiction to Theorem 1. 
        Incidentally, this reasoning also supplies the explanation of the term ``soliton-type field,''
though we left out the question of stability so important for ``true solitons.''
	We remark that other particle-like field solutions without sources are still conceivable, such
as ``breather-type fields,'' but these would be genuinely time-dependent.

	Theorem 2 bears on the quest for electromagnetic self-induction.
	Namely:

\noindent
* \emph{A static electric charge density cannot by itself induce a static magnetic field.}

	This puts to rest interesting recent speculations in \cite{VellozoEtALa, VellozoEtALb}
that the Maxwell--Born--Infeld field equations with a static point charge source would have 
genuinely electromagnetic finite-energy solutions in entire space in which a stationary magnetic
field is induced solely by the static electric field of the point charge.

	The counterpart to the previous item in terms of Theorem 3 is:

\noindent
* \emph{A stationary electric current density alone can only create a magnetostatic field.}

	Theorem 3 also says that perturbation theory works well, whenever it works!

\noindent
* \emph{Whenever the perturbative series in \cite{KieMBIinJMP} converges, and $\rho\equiv 0$ or 
$\jV\equiv\vect{0}$, then no additional, non-perturbative stationary finite-energy  electromagnetic fields exist.}

	Note that this does not rule out a nonperturbative stationary electromagnetic field 
when $\rho\not\equiv 0$ \emph{and} $\jV\not\equiv\vect{0}$ and the perturbative series converges,
or should the perturbative series fail to converge.
	These are interesting open problems!

%%%%%%%%%%%%%%%%%%%%%%%%%%%%%%%%%%%%%%%%%%%%%%%%%%%%%%%%%%%%%%
%%%%%%%%%%%%%%%%%%%%%%%%%%%%%%%%%%%%%%%%%%%%%%%%%%%%%%%%%%%%%%
                \section{Uniqueness results for the Maxwell--Born and other 
	                  nonlinear electromagnetic field equations}
%%%%%%%%%%%%%%%%%%%%%%%%%%%%%%%%%%%%%%%%%%%%%%%%%%%%%%%%%%%%%%
%%%%%%%%%%%%%%%%%%%%%%%%%%%%%%%%%%%%%%%%%%%%%%%%%%%%%%%%%%%%%%

	The main ingredients of our uniqueness proofs are convexity of the Hamiltonian in 
$\BV$ and $\DV$ separately, plus the feature that the space integrals of $\BV\cdot\HV$ and 
of $\EV\cdot\DV$ do not vanish for any nontrivial magnetostatic or electrostatic solution 
involved in the arguments, respectively.
	Whenever these are met, our proofs apply. 
	Of course, convexity of the Hamiltonian in $\DV$ is an immediate consequence of its definition in 
terms of a Legendre-Fenchel transform of the Lagrangian, but 
whether the convexity of the Hamiltonian in $\BV$ holds or not is a different matter; we obtained 
it from the $\BV\leftrightarrow\DV$ symmetry of the Maxwell--Born--Infeld Hamiltonian.
	This symmetry is featured also by some other electromagnetic models, but does not hold in general.
	Also the just mentioned properties of $\BV\cdot\HV$ and $\EV\cdot\DV$ may not automatically hold.

	The nonlinear electromagnetic field theory of Born is an example where all the above holds,
plus more: the Hamiltonian for the Maxwell--Born field equations is jointly strictly convex in 
$\BV$ and $\DV$. 
	This leads to our final theorem which extends the source-free result of Yang \cite{Yisong} to
the general stationary case.
\smallskip

\noindent{\textbf{Theorem 4}.}
\emph{Suppose the set of Maxwell--Born field equations, \refeq{eq:MdotB}--\refeq{eq:FOLIeqHofBD}
without the $\BV\times\DV$ terms, admits a stationary finite-energy solution $(\BV_0,\DV_0)$ 
with stationary sources $\rho$ and $\jV$ satisfying \refeq{eq:MrhojLAW}.
	Then this solution is the unique stationary finite-energy solution for the given $\rho(\sV)$ and 
$\jV(\sV)$.}
\smallskip

\noindent
{\sc{Proof}}: Suppose there were two different stationary solutions with finite field 
energy, say $(\BV_0,\DV_0)$ and $(\BV_1,\DV_1)$, and let $\AV_0$ and $\AV_1$ be the respective 
magnetic vector potentials for $\BV_0$ and $\BV_1$.
	We define a straight interpolating line in $\BV,\DV$ space, given by
$\{(\BV_\lambda,\DV_\lambda): \lambda\in[0,1]\}$ with
$\BV_\lambda = \lambda\BV_1 + (1-\lambda)\BV_0$ 
and 
$\DV_\lambda = \lambda\DV_1 + (1-\lambda)\DV_0$.
	We also define the associated fields
\begin{alignat}{1}
\hskip-.5truecm
{\EV}_{\lambda} 
= 
\frac{{\DV}_{\lambda}}
     {\sqrt{ 1 + \frac{1}{b^2}(|{\BV}_{\lambda}|^2 + |{\DV}_{\lambda}|^2) }}
\quad \mbox{and}\quad
{\HV}_{\lambda} 
 = \label{eq:HandEofBandDlines}
\frac{{\BV}_{\lambda} }
     {\sqrt{ 1 + \frac{1}{b^2}(|{\BV}_{\lambda}|^2 + |{\DV}_{\lambda}|^2) }}
\, .
\end{alignat}
     Accordingly, we also write $\phi_k$ for the electric potentials of the $\EV$ fields, viz.
$\EV_k = -\nab \phi_k$ for $k=1$ and $2$; note that for $\lambda\in(0,1)$ the $\EV_\lambda$ are
generally not gradient fields, though.
     Finally, $\dot{X}$ means derivative of the quantity $X$ w.r.t. $\lambda$.
 
     Now, by hypothesis, both $(\BV_0,\DV_0)$ and $(\BV_1,\DV_1)$ solve the stationary 
Maxwell--Born field equations with the same stationary sources, so for $k=0,1$ we have 
\begin{alignat}{1}
\textstyle
\nab\times\HV_k(\sV)  
&= 
 4\pi \textstyle{\frac{1}{c}} \jV(\sV) \, ,
\label{eq:McurlH}
\end{alignat}
and by the linearity of \refeq{eq:McurlH} in $\HV$ we now have that
\begin{alignat}{1}
\textstyle
\nab\times(\HV_1 - \HV_0)(\sV)  
&= 
\vect{0}
\, .
\label{eq:McurlHdiff}
\end{alignat}
	We take the Euclidean inner product of \refeq{eq:McurlHdiff} with $\AV_1-\AV_0$, integrate over
$\Rset^3$, use an integration by parts to move the $\nab\times$ operator over to the $\AV$'s
(using again that surface integrals at spatial infinity vanish because of the finite-energy 
condition), note that $\nab\times \AV_k =\BV_k$ for $k=1$ and $2$, and obtain
\begin{alignat}{1}
\hskip -.5truecm
0 
= 
\int_{\mathbb R^3} (\AV_1-\AV_0)\cdot \nab\times(\HV_1 - \HV_0) \drm^3s
 =
\int_{\mathbb R^3} (\BV_1-\BV_0)\cdot (\HV_1- \HV_0)  \drm^3s \, .
\label{eq:BHzero}
\end{alignat}
	Similarly, for $k=1$ and $2$ we have
\begin{alignat}{1}
        \nab\cdot\DV_k(\sV)  
&=
        4 \pi \rho (\sV) \, ,
\label{eq:MdivDstat}
\end{alignat}
and the linearity of \refeq{eq:MdivDstat} in $\DV$ gives us
\begin{alignat}{1}
        \nab\cdot(\DV_1-\DV_0)(\sV)  
&=
0
\, .
\label{eq:MdivDstatDIFF}
\end{alignat}
   Multiplying \refeq{eq:MdivDstatDIFF} by $\phi_1-\phi_0$, integrating over
$\Rset^3$, using an integration by parts to move the $\nab$ operator over to the $\phi$'s
(one last time using that surface integrals at spatial infinity vanish because of
the finite-energy condition), noting that $-\nab\phi_k =\EV_k$ for $k=1,2$, we obtain
\begin{alignat}{1}
0 
&= 
\int_{\mathbb R^3} (\phi_1-\phi_0)\nab\cdot(\DV_1 - \DV_0)  \drm^3s
 =
\int_{\mathbb R^3} (\EV_1 - \EV_0)\cdot (\DV_1-\DV_0) \drm^3s \, .
\label{eq:DEzero}
\end{alignat}
   Adding \refeq{eq:BHzero} and \refeq{eq:DEzero} gives
\begin{alignat}{1}
0 = 
\int_{\mathbb R^3} \left[ (\BV_1-\BV_0)\cdot (\HV_1 - \HV_0) +
            (\DV_1-\DV_0)\cdot (\EV_1 - \EV_0)\right](\sV) \drm^3s \,.
\label{eq:BHDEzero}
\end{alignat}
   Now we use that $\BV_1 - \BV_0 = \dot\BV_\lambda$ and $\DV_1 - \DV_0 = \dot\DV_\lambda$, 
while $\HV_1 - \HV_0 = \int_0^1 \dot\HV_{\lambda}\drm{\lambda}$ and 
$\EV_1 - \EV_0 = \int_0^1 \dot\EV_{\lambda}\drm{\lambda}$.
   By the $\lambda$-independence of $\dot\BV_\lambda$ and $\dot\DV_\lambda$, r.h.s.\refeq{eq:BHDEzero} 
becomes
\begin{alignat}{1}
0
= 
\int_{\mathbb R^3} \int_0^1 
\left[ \dot\HV_\lambda \cdot \dot\BV_\lambda +  \dot\EV_\lambda \cdot \dot\DV_\lambda \right](\sV)
\drm\lambda\, \drm^3s \,.
\label{eq:BHDEzeroREDO}
\end{alignat}
 Again by the $\lambda$-independence of $\dot\BV_\lambda$ and $\dot\DV_\lambda$, and by explicit calculation,
\begin{alignat}{1}
\dot{\HV}_{\lambda} \cdot \dot{\BV}_{\lambda} + \dot{\EV}_{\lambda} \cdot \dot{\DV}_{\lambda} 
&=
\textstyle\frac{\drm}{\drm\lambda}
\left({\HV}_{\lambda} \cdot \dot{\BV}_{\lambda} + {\EV}_{\lambda} \cdot \dot{\DV}_{\lambda} \right)\\
&= \label{eq:HlBlElDldotprodID}
 \textstyle b^2\frac{\drm^2}{\drm\lambda^2}
\left({\sqrt{ 1 + \frac{1}{b^2}\big(|{\BV}_{\lambda}|^2+|{\DV}_{\lambda}|^2\big)}} - 1\right).
\end{alignat}
	Note that r.h.s. \refeq{eq:HlBlElDldotprodID} $\geq 0$, with ``$=$'' iff $\BV_\lambda=\vect{0}$ and 
$\DV_\lambda=\vect{0}$ for all $\lambda$.
	Since the solutions are different by hypothesis, we have that somewhere
r.h.s. \refeq{eq:HlBlElDldotprodID} ``$>0$'' strictly, 
but by \refeq{eq:BHDEzeroREDO} and \refeq{eq:HlBlElDldotprodID} this is impossible. 
    Therefore, any stationary finite-energy solution to the electromagnetic Maxwell--Born equations is
unique.~~~QED

\noindent
\textbf{ACKNOWLEDGEMENT:} The work reported here has been funded by  NSF grant DMS-0807705.  
I gratefully acknowledge valuable suggestions by my colleague Shadi Tahvildar-Zadeh.
I also thank the two referees for their interesting comments.

%%%%%%%%%%%%%%%%%%%%%%%%%%%%%%%%%%%%%%%%%%%%%%%%%%%%%%%%%%%%%%%%%%
%%%%%%%%%%%
%%%%%%%%%%% Now the bibliography 
%%%%%%%%%%%
%%%%%%%%%%%%%%%%%%%%%%%%%%%%%%%%%%%%%%%%%%%%%%%%%%%%%%%%%%%%%%%%%%

\end{document}